\documentclass[11pt,a4paper,english,amssymb,nofootinbib,superscriptaddress]{revtex4}
\usepackage{lmodern}

\usepackage[T1]{fontenc}
\usepackage[latin9]{inputenc}
\usepackage{babel}
\usepackage{amsmath}
\usepackage{graphicx}
\usepackage{amssymb}
\usepackage{esint}
\usepackage{color}
\usepackage{soul}

\usepackage[unicode=true, pdfusetitle,
bookmarks=true,bookmarksnumbered=false,bookmarksopen=false,
breaklinks=false,pdfborder={0 0 1},backref=false,colorlinks=false]
{hyperref}
\def\b{\begin{equation}}
\def\e{\end{equation}}

\makeatletter
\@ifundefined{textcolor}{}
{%
 \definecolor{BLACK}{gray}{0}
 \definecolor{WHITE}{gray}{1}
 \definecolor{RED}{rgb}{1,0,0}
 \definecolor{GREEN}{rgb}{0,1,0}
 \definecolor{BLUE}{rgb}{0,0,1}
 \definecolor{CYAN}{cmyk}{1,0,0,0}
 \definecolor{MAGENTA}{cmyk}{0,1,0,0}
 \definecolor{YELLOW}{cmyk}{0,0,1,0}
 }

\usepackage{latexsym}\usepackage{bm}

\makeatother

\makeatother

\begin{document}
\title{ Exotic Massive Gravity:  Causality and a Birkhoff-like Theorem }
\author{Ercan Kilicarslan}

\email{ercan.kilicarslan@usak.edu.tr}

\affiliation{Department of Physics,\\
 Usak University, 64200, Usak, Turkey}

\author{Bayram Tekin}

\email{btekin@metu.edu.tr}

\affiliation{Department of Physics,\\
 Middle East Technical University, 06800, Ankara, Turkey}
\date{\today}

\begin{abstract}
We study the local causality issue via the Shapiro time-delay computations in the on-shell consistent exotic massive gravity in three dimensions. The theory shows time-delay as opposed to time-advance despite having a ghost at the linearized level both for asymptotically flat and anti-de Sitter spacetimes. We also prove a Birkhoff-like theorem: any solution with a hypersurface orthogonal non-null Killing vector field is conformally flat; and find some exact solutions. 
\end{abstract}

\maketitle

\section{Introduction}

Three dimensional spacetime is poor in massless gravitons, yet very rich in massive ones: in addition to the well known  topologically massive gravity (TMG)\cite{DJT}, new massive gravity (NMG) \cite{new_berg}, cubic \cite{Sinha} and Born-Infeld extensions \cite{binm}-which are all based on actions that depend on the metric alone-a new set of theories that lack a purely metric-based action have been found in \cite{mmg,Setare,mmg2}. The field equations of these theories are on-shell consistent, namely they possess a Bianchi identity for the metrics that solve the field equations but not for generic off-shell metrics. These theories are highly restricted 
\cite{emel1,emel2}. Here we shall be interested in the more recent theory, the so called {\it exotic massive gravity} (EMG)  defined in \cite{EMGpa}, and extended and elaborated in various aspects \cite{emg1,emg2,emg3,emg4}. One of the main reasons in searching for new theories in 2+1 dimensions is to try to construct a bulk and boundary unitrary theory which would amount to defining a quantum theory of gravity via the AdS/CFT conjecture. See the summary of the unitarity problem in three dimensional massive gravity theories in \cite{Unitarity}.

Following the discussion of causality of extended gravity theories in \cite{Cam, Kilicarslan} using the Shapiro time-delay computation \cite{Shapiro}, we study the causality of EMG both in asymptotically flat and anti-de Sitter (AdS) spacetimes.  It was realized in \cite{Cam} that the Einstein-Gauss-Bonnet theory is not causal even in the regime when the theory is unitary and moreover addition of finite number of curvature terms in the theory does not solve the problem. The problem has a solution in string theory \cite{Cam,ven} with an infinite tower of massive intermediate states. 
 This naturally prompted the question as to whether three dimensional theories suffer from causality violation. It was shown in \cite{Kilicarslan} that, unlike the higher dimensional theories, causality does not bring in new constraints beyond the unitarity constraints in the then known three dimensional massive gravity theories. Since that work, the EMG theory has emerged and a similar computation in this theory is one of the tasks of this work. 
 
 As discussed in \cite{Kilicarslan}, the usual computation of the Shapiro time delay of a signal is done for a round trip in a black hole background for more than three spacetime dimensions. However, in three dimensions, using the motion of test particles or fields in a shockwave geometry \cite{shock, Dray} created by a massless particle is better suited. This is because we do not know black hole solutions in these theories other than the Banados-Teitelboim-Zanelli(BTZ) black hole \cite{btz} which is only obtained after identifying points of AdS$_3$; and therefore it is not suitable for local causality discussions via the time-delay arguments.

In addition to the causality discussion, we also prove a theorem which is in some sense analogous to the Birkhoff theorem in four dimensions: all solutions of EMG that possess a hypersurface orthogonal non-null Killing vector field is conformally flat. A similar theorem was proven for TMG in \cite{Aliev:1996eh} in a coordinate-independent way and in \cite{Cavaglia:1999si} with explicit coordinates. Here we provide our proof with both methods. Hence to get non-conformally flat solutions, one must introduce twist or rotation.  In addition, we briefly study all solutions of TMG that also solve EMG and give an explicit example which is the squashed AdS$_3$ metric.  

The lay-out of the paper is as follows: In section II and III we study the causality of the theory in flat and AdS spacetimes respectively using the Shapiro time-delay computations. In Section IV we show that all the spacetimes that possess a 
hypersurface orthogonal Killing vector are conformally flat. In Section V, we show that the solutions of TMG are inherited by EMG as long as the coupling parameters of the theories are related in a prescribed way.

\section{Causality in Exotic Massive Gravity}

Here we study the local causality issue in exotic massive gravity via the computation of the Shapiro time-delay or advance.  Time advance would yield a non-causal theory while time-delay would be consistent with a causal one (see \cite{Cam} and \cite{Kilicarslan} for more on this).  For this purpose, let us consider the source-coupled field equations of EMG \cite{EMGpa} 
\begin{equation}
\begin{aligned}
  G_{\mu\nu}+\frac{1}{\mu}C_{\mu\nu}-\frac{1}{m^2}H_{\mu\nu}+\frac{1}{m^4}L_{\mu\nu}=\Theta_{\mu\nu}( T),
  \label{egmg}
  \end{aligned}
\end{equation}
where $ \Theta_{\mu\nu}(T) $ is a complicated "energy-momentum" tensor which is on-shell covariantly conserved and is given explicitly as
\begin{equation}
\Theta_{\mu\nu}(T)=\frac{\lambda}{\mu} \hat T_{\mu\nu} - \frac{\lambda}{m^2} \epsilon_\mu{}^{\rho\sigma}\nabla_\rho \hat T_{\nu\sigma} + 
	\frac{2\lambda}{m^4} \epsilon_\mu{}^{\rho\sigma} \epsilon_\nu{}^{\lambda\tau} 
	C_{\rho\lambda}\hat T_{\sigma\tau} - \frac{\lambda^2}{m^4} \epsilon_\mu{}^{\rho\sigma} \epsilon_\nu{}^{\lambda\tau}  \hat T_{\rho\lambda} \hat T_{\sigma\tau}.
\label{theta}
\end{equation}
here $ \hat{T}_{\mu\nu}=T_{\mu\nu}-\frac{1}{2}g_{\mu\nu}T $ and $T_{\mu\nu}$ is covariantly conserved.  The parameter $\lambda$ appears as a coupling constant between the source and the geometry in a non-trivial, non-homogeneous way as can be seen in the last term of (\ref{theta}).  The tensors on the left-hand side of (\ref{egmg}) are defined as
	\begin{equation}\label{HandL}
C_{\mu\nu}= \epsilon_\mu{}^{\rho\sigma} \nabla_\rho S_{\nu\sigma}\, , \qquad
	H_{\mu\nu}= \epsilon_\mu{}^{\rho\sigma} \nabla_\rho C_{\nu\sigma} \, , \qquad L_{\mu\nu} = \frac12 \epsilon_\mu{}^{\rho\sigma}\epsilon_\nu{}^{\lambda\tau} C_{\rho\lambda}C_{\sigma\tau}\, .
	\end{equation}
where $S_{\mu \nu} := R_{\mu \nu} -\frac{1}{4} g_{\mu \nu} R$ . More explicitly, one has 
\begin{equation}
H_{\mu\nu} = \square S^{\mu \nu} - \nabla^\mu \nabla^\nu S + g^{\mu \nu} S_{\alpha  \beta}^2 -3 S^\mu\,_\alpha S^{\alpha \nu},\hskip .7 cm
 L_{\mu\nu} = \frac{1}{2}  g_{\mu\nu} C_{\rho\sigma}C^{\rho\sigma}- C_{\mu\rho}C^\rho\,_{\nu}\, .
	\end{equation}

The theory (\ref{egmg}), around its {\it flat } vacuum, has two massive spin-2 excitations  with different masses given as 
\begin{equation}
	m_\pm  =  m\bigg(\pm \frac{m}{2\mu}+\sqrt{1+ \frac{m^2}{4\mu^2}}\bigg),
\end{equation}
reflecting its parity non-invariant nature. In the $\mu\to \infty$ limit, the masses coalesce: $m_\pm  =m$ for both helicity +2 and -2 modes. We shall also study this parity-invariant version of the theory.
To analyze the causality issue in this theory, let us consider the shock-wave metric created by a massless point particle moving in a fixed direction, say the $x$-direction. The shock-wave metric written in two null, one spatial coordinates is 
\begin{equation}
ds^2=-du dv+H(u,y) du^2+dy^2,\label{ansatz}
\end{equation}
with the null coordinates defined as $u:=t-x$ and $v:=t+x$ and $y$ is the transverse coordinate. Taking the momentum of the massless source particle to be in the  $+x$ direction, one has $p^\mu=\lvert p\rvert(\delta^\mu_0+\delta^\mu_x)$. Figure 1 depicts the spacetime  region near the source. 
\begin{figure}[h]
\centering
\includegraphics[width=0.5\textwidth]{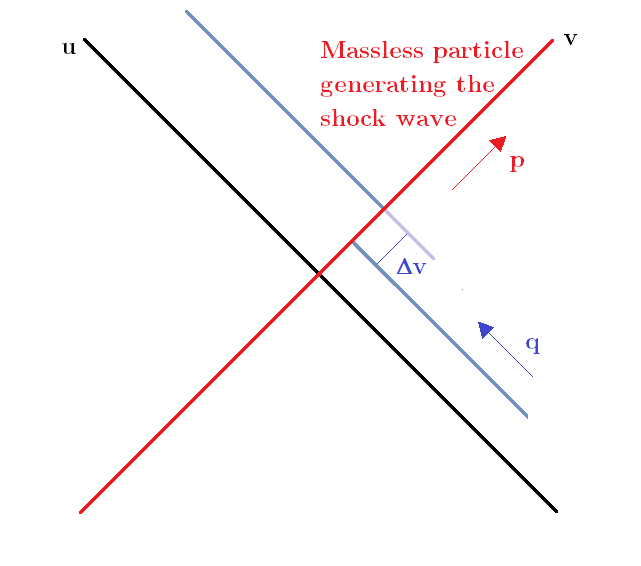}
\caption{Spacetime diagram depicting the shock-wave geometry created by a massless particle with momentum $p$ moving in the $+x$ direction. Another massless test particle with momentum $q$ (blue line) propagating through this geometry, experiences a time-delay when it crosses the constant $u$ line. The time-delay is denoted as a discontinuous jump $\Delta v$ in the null $v$ direction. Time advance would violate causality. }
\label{shapirogeodesicsemg}
\end{figure}
The energy-momentum tensor of such a source has only one non-zero component given as  $T_{uu}=\lvert p\rvert \delta(y)\delta(u)$.
For the shock-wave ansatz (\ref{ansatz}), the scalar curvature $R$ vanishes and the only non-vanishing components of the Ricci, Cotton and $H$ tensors are
\begin{equation}
 R_{uu}=G_{uu}=-\frac{1}{2}\frac{\partial^2}{\partial y^2}H(u,y),\hskip 1 cm C_{uu}=\frac{1}{2}\frac{\partial^3}{\partial y^3}H(u,y),\hskip 1 cm H_{uu}=-\frac{1}{2}\frac{\partial^4}{\partial y^4}H(u,y),
\end{equation}
while the $L$ tensor vanishes identically.
Then the EMG field equations, for the shock-wave metric (\ref{ansatz}), reduce to a single fourth-order differential equation 
\begin{equation}
\begin{aligned}
\bigg (-1+ \frac{1}{\mu}\partial_y+\frac{1}{m^2}\partial_y{^2}\bigg )\partial_y{^2}H(u,y)=2 \lambda\lvert p\rvert \delta(u)\bigg(\frac{\delta(y)}{\mu}+\frac{\delta'(y)}{m^2}\bigg),
\label{eqn}
 \end{aligned}
\end{equation}
where  prime denotes derivative with respect to the transverse coordinate $y$. 
The general solution of the last equation comes with four arbitrary functions  $c_1$, $c_2$, $c_3$ and $c_4$ of the null coordinate $u$, and reads explicitly as 
\begin{equation}
\begin{aligned}
H(u,y)=& \lambda\lvert p \rvert\delta(u)\theta (y)\Bigg\{\big(e^{m_{-} y}-1\big)\bigg(\frac{1}{m_{-}^2}+\frac{\mu+m_{+}}{\mu^2(m_{+}+m_{-})}\bigg)\\&+\frac{\big(e^{-m_{+} y}-1\big)}{m_{+}^2}\bigg(1-\frac{m^2}{\mu(m_{+}+m_{-})}\bigg)+y\bigg(\frac{m_{-}-m_{+}}{m_{-}m_{+}}-\frac{4\mu+m_{+}-m_{-}}{(2\mu-m_{-})(2\mu+m_{+})}\bigg)\Bigg\}\\&+\frac{1}{m^4} \bigg(m^2_{-}e^{-m_{+}y}c_1+m^2_{+}e^{m_{-}y}c_2\bigg)+c_3+c_4 y.
   \end{aligned}
\end{equation}
As noted, this is the most general solution, but one can fix the arbitrary functions by coordinate transformations in such a way that spacetime is asymptotically flat given in Cartesian form far away from the source. But this cannot be done with a single chart for the whole spacetime, so, one can choose the $y > 0$ part to be asymptotically flat in the Cartesian form. For more discussion on this issue, see \cite{Kilicarslan,desershock}.  A careful analysis leads to the following metric profile function with all arbitrary parameters fixed:
\begin{equation}
\begin{aligned}
H(u,y) =&\frac{1}{m_{+}^2}\bigg(1-\frac{m^2}{\mu(m_{+}+m_{-})}\bigg)e^{-m_{+}y}\lambda\lvert p \rvert\delta(u)\theta (y)\\&-y\lvert p \rvert\delta(u)\bigg(\frac{m_{-}-m_{+}}{m_{-}m_{+}}-\frac{4\mu+m_{+}-m_{-}}{(2\mu-m_{-})(2\mu+m_{+})}\bigg)\theta (-y)\\&+\lambda\lvert p \rvert\delta(u)\bigg(\frac{1}{m_{-}^2}+\frac{\mu+m_{+}}{\mu^2(m_{+}+m_{-})}+\frac{1}{m_{+}^2}-\frac{m^2}{\mu(m_{+}+m_{-})m_{+}^2}\bigg)\theta (-y) \\&-\lambda\lvert p \rvert\delta(u)\bigg(\frac{1}{m_{-}^2}+\frac{\mu+m_{+}}{\mu^2(m_{+}+m_{-})}\bigg)e^{m_{-} y}\theta (-y) ~.
\label{metfluc1}
\end{aligned}
\end{equation}
Let us now consider a massless spinless test particle with momentum $q$ traversing the shock-wave created by another massless spinless particle with momentum $p$, with an impact parameter $y=b>0$ as shown in the Figure \ref{shapirogeodesicsemg}. In that region, (\ref{metfluc1}) simplifies and the shock-wave line element is given as 
\begin{equation}
 ds^2=-du\bigg(dv-\frac{1}{m_{+}^2}\bigg(1-\frac{m_{+}-m_{-}}{m_{+}+m_{-}}\bigg)e^{-m_{+}y}\lambda\lvert p \rvert\delta(u)du\bigg)+ dy^2.
\end{equation}
Clearly, as expected, the metric has a discontinuity in the null coordinate $u$ due to distributional nature of the source. This discontinuity can be eliminated by redefining a new null coordinate $v_{new}$ at the impact parameter $b$ as
\begin{equation}
\begin{aligned}
  v_{new}:= v+ \frac{1}{m_{+}^2}\bigg(1-\frac{m_{+}-m_{-}}{m_{+}+m_{-}}\bigg)e^{-m_{+}b}\lambda\lvert p \rvert \theta(u),
 \label{metfluc}
 \end{aligned}
\end{equation}
which gives rise to a time  delay when the particle passes the $u =0$ line as can be explicitly seen from the equivalent expression:
\begin{equation}
\begin{aligned}
 \Delta v =\frac{1}{m_{+}^2}\bigg(1-\frac{m}{\sqrt{m^2+4\mu^2}}\bigg)e^{-m_{+}b}\lambda\lvert p \rvert.
 \label{metflucgemg1}
 \end{aligned}
\end{equation}
Assuming $\lambda >0$,  $\Delta v$ is positive for any value of the impact parameter. It is also important to note that, due to the parity-non invariance of the theory, the test particle experiences a {\it different} time-delay depending on whether it is moving in $+x$ or $-x$ direction. For the impact parameter $y=b<0$, following the similar steps as above, one can find the corresponding time-delay. So for causality, the only constraint is $\lambda >0$. 

Furthermore, in the $\mu\to \infty$ limit and for the choice of $\lambda=m$\footnote{For this choice and limit, we have the matter-coupled field equation \cite{EMGpa} 
 \begin{equation}
   G_{\mu\nu}-\frac{1}{m^2}H_{\mu\nu}+\frac{1}{m^4}L_{\mu\nu}=- \frac{1}{m} \epsilon_\mu{}^{\rho\sigma}\nabla_\rho \hat T_{\nu\sigma} + 
	\frac{2}{m^3} \epsilon_\mu{}^{\rho\sigma} \epsilon_\nu{}^{\lambda\tau} 
	C_{\rho\lambda}\hat T_{\sigma\tau} - \frac{1}{m^2} \epsilon_\mu{}^{\rho\sigma} \epsilon_\nu{}^{\lambda\tau}  \hat T_{\rho\lambda} \hat T_{\sigma\tau},
  \label{emg}
\end{equation}
Note that, in \cite{EMGpa}, a negative sign was erroneously forgotten in the first term of the source. }, which corresponds to the parity-invariant version of EMG theory, the shift in the $v$ coordinate can be written as \footnote{For more details, see Appendix.}
\begin{equation}
\begin{aligned}
 \Delta v =\frac{1}{m}e^{-mb}\lvert p \rvert,
 \label{metflucemg}
 \end{aligned}
\end{equation}
which is again positive for $m >0$. Note that in the opposite limit of  $\mu \rightarrow 0$, the theory boils down to pure Cotton or Chern-Simons theory without a propagating degree of freedom and the metric is locally conformally flat with no interesting dynamics. 
 
\subsection*{Scalar Field in a Shock-Wave}
It pays to reproduce end extend the results of the above computation-done for null geodesics-to fields following \cite{Kilicarslan}. In particular, the results become more transparent for a massless scalar field (still a test field with no back reaction)  propagating in the background shock-wave.  For this purpose, let us consider, the Klein-Gordon (KG) equation for massless real scalar field 
\begin{equation}
\square\phi=0 ,
\label{KG}
\end{equation}
with $\square:=\nabla_\mu\nabla^\mu$, the equation reduces to a non-trivial Partial Differential Equation (PDE) 
\begin{equation}
\partial_u\partial_v\phi+H(u,y)\partial^2_v\phi-\frac{1}{4}\partial^2_y\phi=0 ~,
\end{equation}
whose general solution seems elusive. But for our purposes, this is not needed: all we want is the approximate solution near the shock-wave. In that case, the last term is negligible compared to the others and hence, the massless KG equation becomes 
\begin{equation}
\partial_u\partial_v\phi+H(u,y)\partial^2_v\phi=0 ~,
\label{scalarp}
\end{equation}
which is amenable to a $v$-integration. That integration brings a constant which can be chosen to fit the boundary condition that  in the 
 $v\rightarrow \pm \infty$ limit, the scalar field vanishes.  This reduces the equation to the following first order form
\begin{equation}
\partial_u\phi+H(u,y)\partial_v\phi=0 ~,
\label{scalarp1}
\end{equation}
which admits a solution obtainable by the technique of  separation of variables as 
 $\phi(u,v,y):=U(u)V(v)Y(y)$. Then the solution with a momementum mode $p_v$ reads
\begin{equation}
\phi(u,v,y)=Y(y)U(u_0)V(0)e^{ip_v \big (v-\int^u H(u',y)du'\big )}.
\label{cozum}
\end{equation}
From (\ref{cozum}),
 it is clear that when massless scalar particle/field crosses the shock-wave geometry with an impact parameter $b$, it picks up a Aharonov-Bohm type phase as 
\begin{equation}
\phi(0^{+},v,b)=e^{-ip_{v}\int_{0^{-}}^{0^{+}} du H(u,b)} \phi(0^{-},v,b)=e^{-ip_{v}\Delta v}  \phi(0^{-},v,b) ~,
\end{equation}
here $\Delta v$ is equivalent to the one given in (\ref{metflucgemg1}) obtained via the geodesics computations. Next we extend the discussion to the anti-de Sitter spacetime. 

\section{Causality in Anti-de Sitter Space}

Let us consider the line element of AdS$_{3}$ described in terms of Poincar\'e coordinates as
 \begin{equation}
  ds^2 =\dfrac{\ell^2}{y^2}\bigg(-2dudv+dy^2 \bigg),
 \end{equation}
where  the null $u$,$v$ coordinates were defined in the previous section and take values in the whole real line while  $y\in \mathbb{R}_{+}$. Once again, consider a massless source particle moving in the $+x$ direction in this background; then the resulting shock-wave metric  in the Brinkmann form can be taken as 
\begin{equation} \label{ansatz1}
 ds^{2}=\dfrac{\ell^2}{y^2}\Big(-2dudv+F(u,y)du^2+dy^2\Big),
\end{equation}
with the profile function $F(u,y)$ to be determined below.
The energy-momentum tensor of the point source with the prescribed motion described above at $y_0$ reads
\begin{equation}
T_{uu}=|p|\dfrac{\ell}{y_{0}}\delta(u)\delta(y-y_0). \label{ST}
\end{equation}
The source-coupled field equations of the parity-invariant version EMG take the form~\footnote{ One can consider the generalized version by keeping the Cotton tensor, but the resulting equations are cumbersome without changing the ensuing discussion in a significant way.}   
\begin{equation}
G_{\mu\nu}-\frac{1}{\ell^2}g_{\mu\nu}-\frac{1}{m^2}H_{\mu\nu}+\frac{1}{m^4}L_{\mu\nu}=\Theta_{\mu\nu}(T),
\end{equation}
and for the metric ansatz (\ref{ansatz1}), reduce to a single equation
\begin{equation} \label{egmgeqads}
 \bigg(y^3 \partial_{y}^{3}+2 y^2 \partial_{y}^{2} + m^2 \ell^2 ( -y \partial_{y}+1) \bigg) \partial_{y} F(u,y) = 2 m \ell^2 |p|\delta(u)\bigg(\frac{y \delta(y-y_0)}{y_0}+\frac{y^2\delta'(y-y_0)}{y_0}\bigg),
\end{equation}
whose solution is
\begin{equation}
\begin{aligned}
 F(u,y)=& \frac{\ell^2m\delta(u)\lvert p\rvert}{\ell^2m^2-1}\bigg(\left(\frac{y}{y_0}\right)^{-\ell m+1}+\left(\frac{y}{y_0}\right)^{\ell m+1}-\left(\frac{y}{y_0}\right)^2-1\bigg)\theta(y-y_0) \\&+\frac{\ell^2m\delta(u)\lvert p\rvert}{\ell^2m^2-1}\bigg(c_1\left(\frac{y}{y_0}\right)^{-\ell m+1}+c_2\left(\frac{y}{y_0}\right)^{\ell m+1}
 +c_3\left(\frac{y}{y_0}\right)^2+c_4\bigg),
 \end{aligned} \label{A39}
\end{equation}
where all $c_i$'s depend on $u$. They can be fixed by imposing appropriate boundary conditions we shall do after the following discussion. First let us check the flat space limit of the solution. 

\subsubsection{The Flat Spacetime Limit}

In the flat space limit ($\ell \rightarrow  \infty$), the solution (\ref{A39}) smoothly reduces to the flat space version which we have reproduced in the Appendix for the sake of completeness. To take the limit let us introduce a new coordinate \cite{Kilicarslan}
\begin{equation} \label{change}
 y:=\ell e^{z/\ell},
\end{equation}
in which the AdS$_3$ metric reads 
\begin{equation}
 ds^2=-2 e^{-2z/\ell} dudv+dz^2.
\end{equation}
The flat and AdS shock-wave profile functions are related as   
\begin{equation}
 e^{- 2 z /\ell}F(u,z)=H(u,z).
\end{equation}
Consequently, in the  $\ell \to \infty $ limit, one obtains
\begin{equation}
\begin{aligned}
 H(u,z)=&\dfrac{\delta(u)|p| }{m}\left ( e^{-m(z-z_0)}+ e^{m(z-z_0)}-2\right)\theta(z-z_0)  \\
 &  + \dfrac{\delta(u)|p|}{m}\left(c_1e^{-m(z-z_0)}+ c_2 e^{m(z-z_0)}+c_3+c_4\right), 
 \end{aligned}
\end{equation}
which is the same result as the one found in the flat space analysis. 
\subsubsection{Brown-Henneaux conditions on the AdS$_3$ shock-wave}

 Imposing the Brown-Henneaux (BH) AdS$_3$ boundary conditions \cite{BH} on (\ref{A39}), we can fix the arbitrary functions. BH boundary conditions prescribe decay conditions (as one approaches the connected boundary $y \rightarrow 0$) for the  linearized metric perturbations $h_{\mu \nu }=g_{\mu \nu}-g^{\text{AdS}}_{\mu \nu}$  as 
\begin{equation}
h_{uu}\simeq h_{uv}\simeq h_{vv}\simeq h_{yy}\simeq \mathcal{O}(y^0) , \quad h_{uy}\simeq h_{vy}\simeq \mathcal{O}(y).
\end{equation}
So one demands $F(u,y)\sim\mathcal{O}(y^2)$. The discussion bifurcates depending on the sign of $ 1- \ell m$; for the sake of concreteness, let us assume  $m>{1}/{\ell}$. Then it is clear to see that that one must set $c_1=0$.  Recall that, as one moves to $y \rightarrow \infty$, one approaches the disconnected point boundary of AdS$_3$, we can choose $c_2=-1$  and  $c_3=c_4=1$ to approach AdS$_3$ on that boundary.  Finally, the gauge-fixed shock-wave solution reads
\begin{equation}
\begin{aligned}
F(u,y)=\frac{\ell^2m\delta(u)\lvert p\rvert}{\ell^2m^2-1}\left(\dfrac{y}{y_0}\right)^{1-\ell m}\theta(y-y_0) 
 + \frac{\ell^2m\delta(u)\lvert p\rvert}{\ell^2m^2-1}\left(-\left(\dfrac{y}{y_0}\right)^{1+\ell m}+(\frac{y}{y_0})^2+1\right)\theta(y_0-y). 
\label{denk2}
\end{aligned}
\end{equation}

We can now consider the Shapiro time-delay computation a for massless  scalar field in the AdS shock-wave geometry. In complete analogy with the flat space, one arrives at a Aharonov-Bohm phase and a time-delay given as    
\begin{equation}
 \Delta v=\int_{0^{-}}^{0^{+}}du\ F(u,y) .
\end{equation}
Plugging (\ref{denk2}) in this integral and going to the $z$-coordinates, for 
 $z>z_0$, the time shift can be found to be 
\begin{equation}
 \Delta v=\dfrac{\lvert p\rvert m}{m^2_g}e^{-(z-z_0)(m-\frac{1}{\ell})}, \label{BoundTMG}
\end{equation}
where $m_g$ is the graviton mass given as $m_g^2=m^2- 1/\ell^2$. Observe that, as was shown in flat space analysis, Shapiro time-delay is positive and so causality is not violated in EMG. Note that, if we take the $\ell \to \infty $ limit, we recover the flat space result (\ref {metflucemg}) as expected.

\section{Birkhoff-like Theorem in Exotic Massive Gravity}

For spacetime dimensions $n > 2+1$, the group of spherical symmetry $SO(n-1)$ is non-Abelian. This has a non-trivial consequence on spherically symmetric spacetimes. For example, in four dimensional General Relativity, $SO(3)$ symmetry with three Killing vector fields necessitates a fourth Killing vector field and in particular Ricci-flat spherically symmetric metrics are static  which is the essence of the Birkhoff's theorem (or more properly the Jebsen-Birkhoff theorem \cite{Deser_Jebsen}). On the other hand, in $n=2+1$ spacetime dimensions, which is our case here, the group of ``spherical symmetry'' is $SO(2)$ with a single Killing vector field. This symmetry does not rule out rotations unlike the higher dimensional cases; namely in the $(t,r,\phi)$ coordinates, $g_{t\phi}$ terms need not be zero. So the discussion of the $2+1$ dimensional Birkhoff theorem needs more refinement compared to the four dimensional case. Nevertheless, in topologically massive gravity, a nice theorem was established in a coordinate-independent way in \cite{Aliev:1996eh}; and in local coordinates in \cite{Cavaglia:1999si}.  The essence of the theorem is as follows: {\it in TMG without a cosmological constant, assuming  a {\it{hypersurface}} orthogonal Killing vector field, all solutions are locally flat}. Here, we extend this theorem to the EMG. Unlike the case in TMG, the scalar curvature is not constant and there are higher curvature terms in the equation, so the proof of the analogous theorem is more complicated. 

{\bf Theorem:} Any $2+1$ dimensional spacetime with the cylinder topology ($\Sigma_2\times S^1$) having a non-null hypersurface orthogonal Killing vector field is conformally flat in exotic massive gravity.

{\it Proof:} First let us show this with an explicit construction in local coordinates and later provide the coordinate-free version which is somewhat more involved. Assume local light-cone coordinates  $(u, v,\phi)$ and take the hypersurface orthogonal Killing vector field to be $\xi = \partial_\phi$; then the metric, under the assumptions, can be taken as
\begin{equation}
ds^2=-f(u,v)dudv +g(u,v)^2 d\phi^2,
\label{met}
\end{equation}
here $f(u,v)g(u,v)\ge 0$ is assumed to keep the signature intact. Clearly $\xi$ satisfies the Killing property
\begin{equation}
\nabla_\mu \xi_\nu +  \nabla_\nu \xi_\mu =0,
\end{equation}
and it is also easy to show that it satisfies the hypersurface orthogonality
\begin{equation}
\xi_\mu \nabla_\nu \xi_\sigma +\xi_\nu \nabla_\sigma \xi_\mu+\xi_\sigma \nabla_\mu \xi_\nu =0.
\label{hype}
\end{equation}
 The field equations of the theory in vacuum are
\begin{equation}
\begin{aligned}
	 E_{\mu\nu}&:=G_{\mu\nu}+\Lambda g_{\mu\nu} +\frac{1}{\mu}C_{\mu\nu} - \frac{1}{m^2}H_{\mu\nu} + \frac{1}{m^4} L_{\mu\nu}  \\& 
	 =\mathcal{E}_{\mu\nu} +\frac{1}{\mu}C_{\mu\nu}=0.
\end{aligned}
\label{fieldeq}
	\end{equation}
In the second line we defined
 \begin{equation}
 \mathcal{E}_{\mu\nu}:=G_{\mu\nu}+\Lambda g_{\mu\nu} - \frac{1}{m^2}H_{\mu\nu} + \frac{1}{m^4} L_{\mu\nu}, 
 \end{equation}
 since the crux of the argument is to show that the Cotton tensor will be orthogonal to this tensor. Observe also that we have included a cosmological constant.   
 For the metric (\ref{met}), one can compute $\mathcal{E}^\mu{}_\nu$ and  $C^\mu\,_\nu$ and depict the non-zero parts as~\footnote{It is advantageous to  study the $(1,1)$ tensor form of the field equations instead of $(0,2)$ tensor form.} 
\begin{equation}
\mathcal{E}^\mu\,_\nu=\begin{pmatrix}
X_1 & X_2& 0 \\ 
X_3& X_4 & 0\\ 
0 & 0 &X_5
\end{pmatrix},
\label{matris1}
\end{equation}
and
\begin{equation}
C^\mu\,_\nu=\begin{pmatrix}
0 & 0 & Y_1\\ 
0 & 0  & Y_2\\ 
Y_3 & Y_4 & 0 
\end{pmatrix},
\label{matris2}
\end{equation}
where  $X_i$ and $Y_i$ are complicated functions of $f(u,v)$ and $g(u,v)$ and their derivatives which we shall not write here explicitly. The crucial observation is that the matrices $\mathcal{E}^\mu\,_\nu$ and $C^\mu\,_\nu$ are orthogonal to each other and hence, assuming the field equations (\ref{fieldeq}), they must separately vanish. Vanishing of the Cotton tensor is the necessary and the sufficient condition for a 3D metric to be conformally flat. Therefore, the theorem follows. To obtain conformally non-flat solutions, one must introduce twist, or the Killing vector field should not be hypersurface orthogonal. 

We can verify the above result in a coordinate-free way following the computation in the TMG case given in \cite{Aliev:1996eh}. The hypersurface orthogonal Killing vector field defines a parallel direction and two perpendicular directions which yield a natural splitting of the field equations. For this purpose, let us define the orthogonal-projector $\perp$ as
\begin{equation}
\perp^\mu_\nu := \delta^\mu_\nu - \frac{ \xi^\mu \xi_\nu}{\xi^2},
\end{equation}
 where $\xi^2 := g_{\mu \nu} \xi^\mu \xi^\nu \ne 0$. So clearly $\perp^\mu_\nu  \xi^\nu =0$.
In what follows we shall denote the component of a tensor in the direction parallel to $\xi$ as $T_\xi$ and perpendicular to $\xi$ as $T_\perp$.  Let us first show the following
\begin{equation}
C^\xi\,_\xi =0, \hskip 1 cm   \mathcal{E}^\xi\,_\xi \ne 0,
\label{eqBir}
\end{equation}
which are the bottom far right corners in (\ref{matris2}) and (\ref{matris1}). By definition 
\begin{equation}
C^\xi\,_\xi = C^\mu\,_\nu \xi_\mu \xi^\nu  = \xi_\mu \xi^\nu \eta^{\mu \alpha \beta}\nabla_{\alpha} \left (R_{\beta \nu} - \frac{1}{4} g_{\beta \nu} R \right) = \xi_\mu \xi^\nu \eta^{\mu \alpha \beta}\nabla_{\alpha}R_{\beta \nu}, 
\end{equation}
where the scalar curvature term $R$ dropped due to symmetry, {\it not} due to $R$ being a constant as in TMG. In fact $R$ is not assumed to be a constant. Pulling out the covariant derivative, one has 
\begin{equation}
C^\xi\,_\xi = \nabla_{\alpha}\left (\xi_\mu \xi^\nu \eta^{\mu \alpha \beta}R_{\beta \nu} \right) -  \eta^{\mu \alpha \beta}R_{\beta \nu}\bigg ( \xi_\mu \nabla_{\alpha} \xi^\nu + \xi^\nu \nabla_{\alpha} \xi_\mu \bigg). 
\label{pr2}
\end{equation}
To proceed we need some identities for the assumed $\xi$ derived in \cite{Aliev:1996eh}. By taking the derivative $\nabla^\nu$ of (\ref{hype}), one finds the following identity
\begin{equation}
 \xi_\mu R_{\alpha \nu}\xi^\nu = \xi_\alpha R_{\mu \nu}\xi^\nu,
\label{guzid}
\end{equation}
which basically says that $\xi$ can be used to barter an index of the once-contracted Ricci tensor. This identity kills the first term in (\ref{pr2}). For the second part we need the following identity which can be obtained by contracting (\ref{hype}) with $\xi^\mu$:
\begin{equation}
\nabla_\mu \xi_\nu = \frac{1}{2} \left( \xi_\nu \partial_\mu \log |\xi^2| - \xi_\mu \partial_\nu \log |\xi^2| \right).
\end{equation}
Making use of this identity in the second part of (\ref{pr2}), one has 
\begin{equation} 
\eta^{\mu \alpha \beta}R_{\beta \nu}\bigg ( \xi_\mu \nabla_{\alpha} \xi^\nu + \xi^\nu \nabla_{\alpha} \xi_\mu \bigg) = \frac{3}{2}\eta^{\mu \alpha \beta}R_\beta\,^\nu \xi_\mu \xi_\nu \partial_\alpha \log |\xi^2|, 
\end{equation}
which vanishes upon use of (\ref{guzid}). Hence for non-null hypersurface orthogonal Killing vector one has  $C^\xi\,_\xi \equiv 0$. On the other hand one has 
\begin{equation}
\mathcal{E}^\xi\,_\xi = R^\mu\,_\nu \xi_\mu \xi^\nu +(\Lambda - \frac{R}{2})\xi^2  - 
\frac{1}{m^2}H^\mu\,_\nu \xi_\mu \xi^\nu+ \frac{1}{m^4} L^\mu\,_\nu \xi_\mu \xi^\nu.
\end{equation}
It is not difficult to see that there is no reason for this expression to vanish identically for example one has $ R^\xi\,_\xi = - \xi^\mu \square \xi_\mu \ne 0 $. Therefore (\ref{eqBir}) is proven.

Let us now prove the following 
\begin{equation}
\mathcal{E}^\xi\,_\perp \equiv 0, \hskip 1 cm  C^\xi\,_\perp \ne 0.
\label{eqBir2}
\end{equation}
We have, by definition
\begin{equation}
\mathcal{E}^\xi\,_\perp = R^\mu\,_\nu \xi_\mu \perp^\nu_\alpha  - 
\frac{1}{m^2}H^\mu\,_\nu \xi_\mu\perp^\nu_\alpha+ \frac{1}{m^4} L^\mu\,_\nu \xi_\mu \perp^\nu_\alpha.
\label{geneq}
\end{equation}
Let us study this term by term as each term must vanish independently if the expression is expected to vanish identically due to the inhomogeneity of the expression in the mass parameter $m$. The first term is easy:
\begin{equation}
R^\xi\,_\perp :=R^\mu\,_\nu \xi_\mu \perp^\nu_\alpha = R^\mu\,_\nu \xi_\mu \left ( \delta^\nu_\alpha - \frac{ \xi^\nu \xi_\alpha}{\xi^2} \right) = R^\mu\,_\alpha \xi_\mu - R^\mu\,_\nu \xi_\mu \frac{ \xi^\nu \xi_\alpha}{\xi^2} =0,
\end{equation}
where we used the index-bartering identity (\ref{guzid}) in the second term. Similarly the second term in (\ref{geneq}) reads
\begin{equation}
\begin{aligned}
H^\xi\,_\perp :&=H^\mu\,_\nu \xi_\mu \perp^\nu_\alpha  
=\frac{1}{2}\eta^{\nu\mu}\,\,_\alpha\nabla_\nu\bigg(\xi^\sigma C_{\mu\sigma}\bigg)-\frac{1}{2}\eta^{\nu\mu\lambda}\frac{\xi_\lambda\xi^\alpha}{\xi^2}\nabla_\nu\bigg(\xi^\sigma C_{\mu\sigma}\bigg),
\label{Heq}
\end{aligned}
\end{equation}
where we used the fact that $\xi$ is a Killing vector yielding
\begin{equation}
{\cal{L}}_\xi C_{\mu\nu}= 0,
\end{equation}
which can be used to show the following relation
\begin{equation}
\xi_\sigma H^\lambda\,_\sigma=\frac{1}{2}\eta^{\nu\mu\lambda}\nabla_\nu\bigg(\xi^\sigma C_{\mu\sigma}\bigg). 
\end{equation}
It is clear that the $\alpha$ index in (\ref{Heq}) must be $\perp$, hence one has
\begin{equation}
H^\xi\,_\perp =\frac{1}{2}\eta^{\xi\perp}\,\,_\perp\nabla_\xi C^\xi\,_\perp.
\end{equation}
To show that this vanishes, we need to following $\nabla_\mu C^\mu\,_\nu=0$ which yields $\nabla_\xi C^\xi\,_\perp=0$ since $C^\perp\,_\perp=0$. Let us show in fact that $C^\perp\,_\perp=0$ even when the scalar curvature is not constant:
\begin{equation}
\begin{aligned}
C^\perp\,_\perp:&=C^\mu\,_\nu \perp^\nu\,_\alpha \perp^\beta\,_\mu\\&
=\frac{1}{\xi^2}\bigg(\eta^{\beta\sigma\rho}\nabla_\sigma R_{\rho\alpha}\xi^\mu\xi_\mu-\eta^{\mu\sigma\rho}\nabla_\sigma R_{\rho\alpha}\xi^\beta\xi_\mu-\eta^{\beta\sigma\rho}\nabla_\sigma R_{\rho\nu}\xi^\nu\xi_\alpha
\bigg)\\&
+\frac{\xi_\mu \nabla_\sigma R }{4\xi^2}\bigg(-\eta^{\beta\sigma}\,\,_\alpha\xi^\mu+\eta^{\mu\sigma}\,\,_\alpha\xi^\beta+\eta^{\beta\sigma\mu}\xi_\alpha
\bigg).
\label{cdikdik}
\end{aligned}
\end{equation}
To see that this vanishes requires a couple steps and the use of the three dimensional identity
\begin{equation}
\eta^{\lambda\nu\alpha}\xi^\rho=g^{\lambda\rho}\eta^{\beta\nu\alpha}\xi_\beta+g^{\nu\rho}\eta^{\lambda\beta\alpha}\xi_\beta+g^{\alpha\rho}\eta^{\lambda\nu\beta}\xi_\beta,
\end{equation}
together with the Killing property $ {\cal{L}}_\xi R_{\mu\nu}= 0$ and (\ref{guzid}). After making use of these, one can show that the first and second lines of (\ref{cdikdik}) vanish identically separately. 

Similarly the third term in (\ref{geneq}) reads
\begin{equation}
\begin{aligned}
L^\xi\,_\perp :=L^\mu\,_\nu \xi_\mu \perp^\nu_\alpha&=\bigg(\frac{1}{2}\delta^\mu_\nu C_{\rho\sigma}^2-C^\mu\,_\sigma C^\sigma\,_\nu\bigg)\xi_\mu \perp^\nu_\alpha \\&
=-C^\mu\,_\sigma C^\sigma\,_\nu \xi_\mu \perp^\nu_\alpha\\&
=-C^\xi\,_\perp C^\perp\,_\xi \perp^\xi_\alpha
=0.
\end{aligned}
\end{equation}
where we used the fact that $C^\xi\,_\xi=0$ and $C^\perp\,_\perp=0$. One can also show that $\mathcal{E}^\perp\,_\perp\neq 0$; hence the theorem follows and one must introduce twist to find conformally non-flat solutions. 

\section{All Solutions of TMG solving Exotic Massive Gravity}	
Field equations of  EMG are highly complicated, but it is clear that all Einstein metrics solve these equations. To move beyond Einstein metrics, let us consider {\it{all}} solutions of TMG that solve  EMG. As the solutions of TMG are compiled in a nice paper \cite{Pope}, we shall not go into an extended discussion here, but just find the conditions that are needed to carry the TMG to solutions to the current theory.
 Let us assume that the metric $g_{\mu\nu}$ solves TMG whose topological mass is $1/a$; hence it satisfies the following equations
\begin{equation}
C_{\mu\nu}=a\tilde{R}_{\mu\nu}, \hskip 1.cm R=k,
\label{tmg3}
\end{equation}
where $a$ and $k$ are constant and $\tilde{R}$ is the traceless Ricci tensor defined as $\tilde{R}_{\mu\nu}=R_{\mu\nu}-\frac{1}{3}g_{\mu\nu}R$. To search for solutions of (\ref{fieldeq}), we assume that (\ref{tmg3}) also holds.  Therefore, we are searching for constant scalar curvature solutions. The trace part of the field equations (\ref{fieldeq}) is	
\begin{equation}
-\frac{R}{2}+3\Lambda+ \frac{1}{2m^4} C_{\mu\nu}C^{\mu\nu} =0\, ,
\label{treqn}
\end{equation}
and the traceless part is
\begin{equation}
\tilde{R}_{\mu\nu}+\frac{1}{\mu}C_{\mu\nu}- \frac{1}{m^2}H_{\mu\nu}+\frac{1}{m^4}\tilde{L}_{\mu\nu}=0\, ,
\label{tracelesseq}
\end{equation}
where $\tilde{L}_{\mu\nu}=\frac{1}{3}g_{\mu\nu}C^2_{\rho\sigma}-C_{\mu\rho}C^{\rho}_{\nu}$. Let us define the following curvature invariants
\begin{equation}
I=\tilde{R}^\mu\,_\nu\tilde{R}^\nu\,_\mu, \hskip 1.cm J=\tilde{R}^\mu\,_\alpha\tilde{R}^\alpha\,_\beta \tilde{R}^\beta\,_\mu,
\end{equation}
which are relevant to the classification of the solutions (see \cite{Pope, Gurses_exact} for more on this).
Contracting (\ref{tracelesseq}) with $\tilde{R}_{\mu\nu}$ and making use of the TMG equation, one arrives at 
\begin{equation}
I\bigg(1+\frac{a}{\mu}-\frac{a^2}{m^2}\bigg)-\frac{a^2}{m^4}J =0,
\end{equation}
where $I=\frac{2m^4}{a^2}\bigg(\frac{R}{2}-3\Lambda\bigg)$ which comes from (\ref{treqn}). Then plugging this to the previous equation, one has
\begin{equation}
J =\frac{2m^8}{a^4}\bigg(1+\frac{a}{\mu}-\frac{a^2}{m^2}\bigg)\bigg(\frac{R}{2}-3\Lambda\bigg).
\end{equation}
So the solutions of TMG also solve EMG as long as these $I$ and $J$ equations are satisfied. Let us give an explicit example which is called the time-like squashed $AdS_3$
\begin{equation}
ds^2 = \frac{\lambda^2-4}{ 2 R} \bigg ( - \lambda^2 (d\tau + \cosh \theta d \phi)^2 + d \theta^2 + \sinh^2 \theta d\phi^2\bigg),
\label{timelike}
\end{equation}
with the squashing parameter $\lambda$ (not to be confused with the coupling constant of the earlier sections) and the constant scalar curvature of this metric is $R$. 
The metric (\ref{timelike}) is a solution to EMG if $\lambda$ and $R$ have real solutions in terms of $\mu$ and $\Lambda$ and $m$ as given in the following equations
\begin{equation}
\begin{aligned}
&\mu=\frac{ \sqrt{18R}  
   \left(\lambda ^2-4\right)^{3/2}
   m^4\lambda }{2 \left(\lambda
   ^2-4\right)^2 m^4-9 \lambda ^2
   \left(\lambda ^2-4\right) m^2 R+12
   \lambda ^2 \left(\lambda
   ^2-1\right) R^2}\,,\\&
\Lambda=\frac{R \left(\left(\lambda
   ^2-4\right)^3 m^4-24 \lambda ^2
   \left(\lambda ^2-1\right)^2
   R^2\right)}{6 \left(\lambda
   ^2-4\right)^3 m^4}.
\end{aligned}
\end{equation}
To search for solutions which are more general than the ones that solve TMG, one can resort to the method developed in \cite{gurses_Killing}. 

\section{Conclusions}

In \cite{Cam}, rather unexpectedly, the Einstein-Gauss-Bonnet theory was shown to violate causality for any sign of the Gauss-Bonnet coupling constant. That seems to be a major blow for effective gravity theories. But luckily, string theory with an infinite tower of intermediate states can solve the problem \cite{ven}. Interestingly, despite having their own  problems, various  three dimensional massive gravity theories were shown to not suffer from the causality violations; since  a detailed study shows that the conditions coming from causality are not in conflict with the ones coming from unitarity. These theories were discussed in \cite{Kilicarslan} save the recently constructed EMG theory which has not been hitherto studied along these lines. Here we discussed the issue of local causality in EMG in asymptotically flat and AdS spacetimes using the Shapiro time-delay computation for massless test particles and scalar fields in a shock-wave geometry created by a massless source. Despite having a ghost, there is a time-delay for any impact parameter between the source and the test field instead of a time-advance hence causality is not violated. In addition, we have studied some exact solutions in the theory and proved that all solutions with a hypersurface orthogonal Killing vector field are conformally flat. To go beyond conformally flat solutions, rotation must be introduced.  

\section*{Appendix: Some details of causality in the parity-invariant version of  EMG}
The field equations of parity-invariant version of EMG  are \cite{EMGpa}
\begin{equation}
\begin{aligned}
G_{\mu\nu}-\frac{1}{m^2}H_{\mu\nu}+\frac{1}{m^4}L_{\mu\nu}=\Theta_{\mu\nu}(T),
\label{egmg1}
\end{aligned}
\end{equation}
where $ \Theta_{\mu\nu}(T) $ is energy momentum tensor and it is given as
\begin{equation}
\Theta_{\mu\nu}(T)=  - \frac{1}{m} \epsilon_\mu{}^{\rho\sigma}\nabla_\rho \hat T_{\nu\sigma} + 
\frac{2}{m^3} \epsilon_\mu{}^{\rho\sigma} \epsilon_\nu{}^{\lambda\tau} 
C_{\rho\lambda}\hat T_{\sigma\tau} - \frac{1}{m^2} \epsilon_\mu{}^{\rho\sigma} \epsilon_\nu{}^{\lambda\tau}  \hat T_{\rho\lambda} \hat T_{\sigma\tau},
\end{equation}
here $ \hat{T}_{\mu\nu}=T_{\mu\nu}-\frac{1}{2}g_{\mu\nu}T $.
For the shock-wave metric, field equations (\ref{egmg1}) reduce to a single equation
\begin{equation}
\left (-1 +\frac{1}{m^2} \partial_y{^2}\right) \partial_y{^2}H(u,y)=2\lvert p \rvert \delta(y)\delta(u),
\label{eqnnmg}
\end{equation}
whose general solution can be found to be  
\begin{equation}
\begin{aligned}
H(u,y) =\frac{ \lvert p \rvert\delta(u)}{ m}\bigg(e^ {-my}+e^ {my}-2\bigg)\theta(y)+\frac{1}{m}\bigg(c_1e^ {my}-c_2e^ {-my}\bigg)+c_3,
\end{aligned}
\end{equation}
with $c_i$ that depend on the null coordinate $u$. 
By gauge fixing as was done in the text for the more general theory, the solution takes the following form
\begin{equation}
\begin{aligned}
H(u,y) =\frac{ \lvert p \rvert\delta(u)}{ m}e^ {-my}\theta(y)+\frac{\lvert p \rvert\delta(u)}{m}\bigg(-e^ {my}+2\bigg)\theta(-y).
\end{aligned}
\end{equation}
Finally, using the discontinuity in this profile function, one can calculate the time-delay of a signal passing at an impact parameter $b$ as 
\begin{equation}
\begin{aligned}
\triangle v =\frac{ \lvert p \rvert}{ m}e^ {-m\lvert b\rvert},
\label{EMGtimedelay}
\end{aligned}
\end{equation}
which is positive and matches (\ref{metflucemg}) for $m >0$.

\end{document}